\documentclass[aps,prx,reprint,superscriptaddress,nobibnotes,longbibliography]{revtex4-2} 

\usepackage{graphicx,amsmath,amssymb}
\usepackage{xspace}
\usepackage{siunitx}
\usepackage[colorlinks]{hyperref}
\usepackage{textcomp}
\usepackage{xcolor}
\usepackage{soul} 
\usepackage{physics}
\usepackage{times}

\newcommand{\Ersq}{E_R}

\begin{document}

\title{Quantum gas microscopy of three-flavor Hubbard systems} 

\author{Jirayu Mongkolkiattichai}
\thanks{Present address: National Astronomical Research Institute of Thailand,\\260 Moo 4, Donkaew, Mae Rim, Chiang Mai, 50180, Thailand.}
\author{Liyu Liu}

\affiliation{Department of Physics, University of Virginia, Charlottesville, Virginia 22904, USA}

\author{Sohail Dasgupta}

\affiliation{Department of Physics and Astronomy, Rice University, Houston, Texas, USA}

\author{Kaden R. A. Hazzard}

\affiliation{Department of Physics and Astronomy, Rice University, Houston, Texas, USA}
\affiliation{Smalley-Curl Institute, Rice University, Houston, Texas, USA}

\author{Peter Schauss}

\affiliation{Department of Physics, University of Virginia, Charlottesville, Virginia 22904, USA}
\affiliation{Institut für Quantenphysik, Universit\"at Hamburg, Luruper Chaussee 149, 22761 Hamburg, Germany}
\email[Corresponding author: ]{peter.schauss@uni-hamburg.de}

\begin{abstract}
Hubbard systems are paradigmatic realizations of strongly correlated many-body systems. Introducing additional species breaks the SU(2) symmetry of the Hubbard model and leads to a wide variety of novel exotic quantum phases. Three-component fermionic systems are at the heart of model systems for quantum chromodynamics where the three components reflect the three flavors. Here, we extend quantum gas microscopy to three-flavor Fermi lattice gases in the Hubbard regime. Relying on site- and flavor-resolved detection, we study the phase diagram of the three-flavor Hubbard model and find signatures of flavor-selective localization and selective pairing at temperatures down to the tunneling energy scale. Our measurements are compared with numerical linked-cluster expansion calculations. Further increase of phase space density may enable the observation of a novel pair Mott phase at half filling, and shows a path towards the study of color superfluidity and other aspects of quantum chromodynamics.
\end{abstract}

\maketitle

Multi-component Fermi gases as a generalization of two-component gases have attracted considerable interest as model systems for the formation of novel quantum phases including color superfluidity \cite{Cherng2007,Rapp2007}.
In Fermi gases of three components, the components resemble the three colors of fermions in quark matter \cite{Wiese2014,Banuls2020,Chetcuti2023}. With ultracold atoms, the different components can be represented by multiple nuclear spin states in the ground state. SU($N$) systems and three-component Fermi systems have been studied in free-space \cite{Pagano2014,Ozawa2018,Sonderhouse2020,Sonderhouse2020,Song2020}, and have been found to be surprisingly long-lived despite the presence of three-body losses.

Fermi-Hubbard models have been introduced as a model system to describe the physics of strongly correlated electronic systems \cite{Hubbard1963} and to capture essential features of high-temperature superconductivity \cite{Esslinger2010,Dagotto1994}.
But variations of Hubbard models can simulate a wide variety of physical models, along the ideas of lattice quantum chromodynamics \cite{Banerjee2013,Tajima2022,Ciavarella2023, Xu2023_trion}.
Recently, there has been an increased interest in extensions of the standard Hubbard model. This is in part driven by the observation that the square-lattice SU(2) Hubbard model has exceptional coincidental properties rarely realized in real materials. These include particle-hole symmetry at half filling and perfect nesting. The Hubbard model can be displaced from this exceptional point by interactions beyond on-site  \cite{Baier2016,Su2023}, modification of the lattice geometry that lead to kinetic frustration \cite{Mongkolkiattichai2023,Xu2023,Prichard2024,Lebrat2024} or by introducing multiple species. Systems with more than two components host a wider variety of ordering in optical lattices \cite{Sotnikov2014,Yanatori2016_SUN,Hafez-Torbati2019,Ibarra-Garc2023}. 
For the past decade, the study of SU($N$) Fermi-Hubbard models using alkaline-earth-like atoms has attracted broad interest and these systems are expected to show exotic spin ordering and modifications compared to SU(2) Hubbard physics. Tremendous experimental progress has been made towards observing these effects by realizing SU(6) Mott insulators in the 3d cubic lattice \cite{Taie2012,Tusi2022}, measuring the equation of state of SU($N$) models \cite{Hofrichter2016,Pasqualetti2024} and detecting antiferromagnetic correlations in the SU(6) Fermi-Hubbard model in 1d, 2d square, and 3d cubic lattice geometries \cite{Taie2022}.

In this work, we image three-flavor Fermi lattice systems with equal flavor-populations but broken SU(3) symmetry with single-site resolution in a flavor-resolved quantum gas microscope. We study those Fermi lattice systems with unequal tunable interactions over a wide parameter regime. The flavor-resolved imaging enables us to investigate the interplay among pairs of flavors. We focus on flavor-balanced mixtures with equal atom number in all three flavors and choose regimes where three-body losses are suppressed. Depending on the interaction strengths, we observe the onset of the three-flavor Mott insulating state, selective pairing of two flavors, and competing pairing. We also see both ferromagnetic and antiferromagnetic correlations in the same system between different flavors. Three-flavor systems with unequal interaction strength have been studied numerically and it has been predicted that they host a variety of quantum phases, in particular different classes of Mott insulating states, depending on the interaction strengths \cite{Miyatake2010,Inaba2010,Inaba2013}. In this study, we perform quantitative comparisons to numerical calculations \cite{IbarraGarca2021,Feng2023,Ibarra-Garc2023} and directly observe signatures of the predicted phases. Our observations rely on site-resolved and flavor-resolved images of three-component Hubbard systems, enabling the direct detection of intra- and inter-flavor correlations, including pairing between any two flavors.

For many atomic species, three-component mixtures lead to high loss rates but $^6$Li is an exception. The stability of three-component mixtures of ultracold fermionic $^6$Li was characterized by the three-body loss rate coefficient at various fields for equal populations \cite{Ottenstein2008,Huckans2009} and population imbalance \cite{Schumacher2023}. In addition, the direct measurement of binding energy by radio-frequency spectroscopy was used to probe the existence of Efimov trimers in the system \cite{Lompe2010}.

\begin{figure*}
\includegraphics[width=0.7\linewidth]{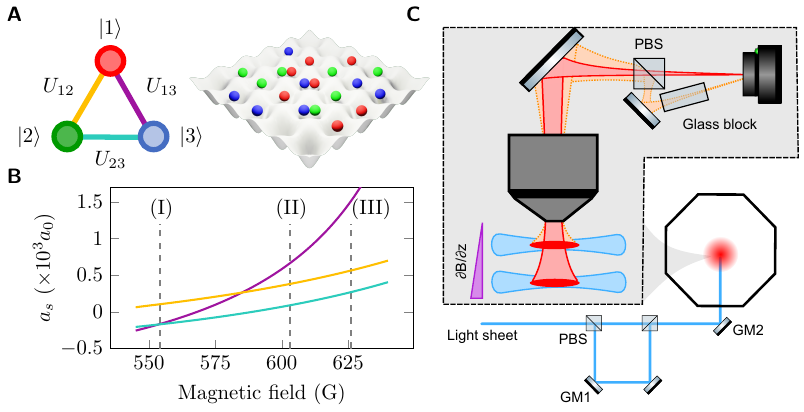}
\caption{{\bfseries Quantum gas microscopy of three-flavor Fermi gases.} {\bfseries(A)} Definition of flavors and interactions between pairs of flavors (left). Illustration of a three-flavor Fermi lattice gas (right). {\bfseries(B)} Tunability of the three pair interactions, $U_{13}$ (violet), $U_{12}$ (orange), and $U_{23}$ (cyan). (I)-(III) represent different interaction strengths at magnetic fields of 555.9(1), 603.3(1), and 625.9(3)~G chosen for detailed measurements. {\bfseries(C)} Detection scheme for three-flavor Fermi lattice gases. The elliptical Gaussian beam (light sheet) is equally split into two beams “LS1” (short path) and “LS2” (long path) using a polarizing beam splitter (PBS). To prevent an expansion of the beam in the long path, a 1:1 telescope is added (not shown), making the shape of LS2 identical to LS1. Both light sheets are combined with another PBS. To individually control the movement of light sheets in the vertical direction, we attach mirrors to galvanometers (GM1, GM2) \cite{SuppOnline}. Fluorescence light is collected through a high-resolution objective and both layers are simultaneously focused on the camera. A glass block with a diameter of 25~mm and a length of 63.5~mm is used to compensate for the focus shift between the two layers. The relative angle between the two paths is approximately 12$^\circ$ at the camera.}
\label{fig: setup}
\end{figure*}

To realize the three flavors in a Fermi system in the experiment, we work with the three lowest hyperfine states of $^6$Li defined by $\ket{1}=\ket{F=1/2,m_F=1/2}$, $\ket{2}=\ket{F=1/2,m_F=-1/2}$, and $\ket{3}=\ket{F=3/2,m_F=-3/2}$ where $F$ and $m_F$ are the hyperfine and magnetic quantum numbers. We label the three flavors, $\ket{1}$ (red), $\ket{2}$ (green), and $\ket{3}$ (blue) and represent their pair interactions by $U_{12}$ (orange), $U_{23}$ (cyan), and $U_{13}$ (violet) (Fig.~\ref{fig: setup}A). The nature of Feshbach resonances in $^6$Li enables us to tune the relative interaction strengths (Fig.~\ref{fig: setup}B) from all repulsive interactions (II,III) to one repulsive and two attractive interactions (I). In addition, by varying the lattice depth a wide range interaction strengths in units of the tunneling strength are accessible.

Our system is well described by the three-flavor Fermi-Hubbard model in a two-dimensional square lattice. The Hamiltonian is given by
\begin{widetext}
\begin{equation}
\hat{\mathcal{H}}=-t \sum_{\langle \mathbf{rr^\prime}\rangle} \sum_{\alpha=1}^3 (\hat{c}_{\mathbf{r} \alpha}^{\dagger} \hat{c}_{\mathbf{r^\prime} \alpha}+\hat{c}_{\mathbf{r^\prime \alpha}}^{\dagger} \hat{c}_{\mathbf{r} \alpha})
-\sum_\mathbf{r} \sum_{\alpha=1}^3 \mu_\alpha(\mathbf{r}) \hat{n}_{\mathbf{r} \alpha}
+\sum_\mathbf{r} \sum_{\alpha < \beta} U_{\alpha \beta} \hat{n}_{\mathbf{r} \alpha} \hat{n}_{\mathbf{r} \beta},
\end{equation}
\end{widetext}
where $\alpha,\beta\in\{1,2,3\}$ represent the fermionic flavors that correspond to the three lowest hyperfine states of $^6$Li, $t$ is the tunneling strength between nearest-neighbor lattice sites, $U_{\alpha\beta}$ is the on-site interaction between flavors $\alpha$ and $\beta$, $\hat{c}_{i \alpha}(\hat{c}_{j \alpha}^{\dagger})$ is the annihilation (creation) operator for a fermion of flavor $\alpha$ on site $\mathbf{r}$, $\hat{n}_{\mathbf{r},\alpha}=\hat{c}_{\mathbf{r}\alpha}^{\dagger} \hat{c}_{\mathbf{r} \alpha}$ is the number operator, and $\mu_{\alpha}(\mathbf{r})$ is the chemical potential for flavor $\alpha$. Our lattice potential has a harmonic confinement and we apply the local density approximation (LDA) to map the local observables to the homogeneous case: $\mu_\alpha(\mathbf{r})=\mu_{0,\alpha}-(1/2)m\omega^2r^2$ where $\omega$ is the lattice confinement, $m$ the atomic mass of $^6$Li, and $\mu_{0,\alpha}$ the chemical potential at the trap center for the flavor $\alpha$.

\begin{figure*}
\includegraphics[width=0.6\linewidth]{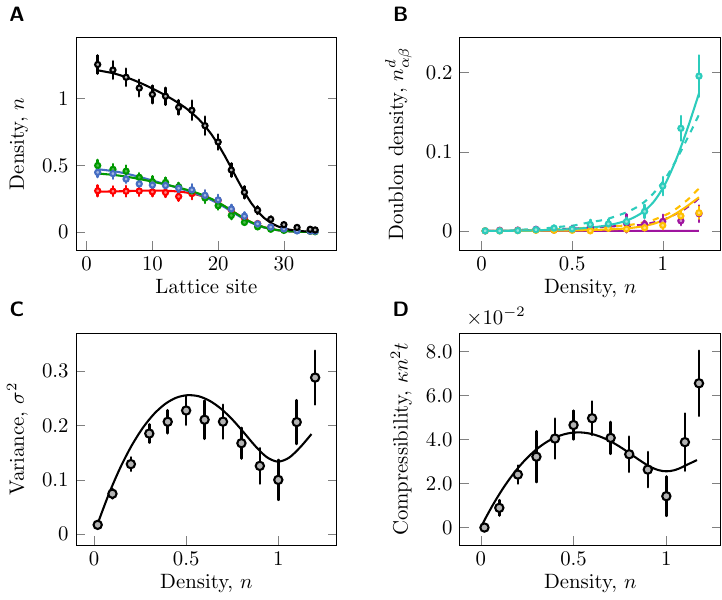}
\caption{{\bfseries Three-flavor Mott insulators.} {\bfseries(A)} Density as a function of radius from the trap center for $\ket{1}$ (red), $\ket{2}$ (green), and $\ket{3}$ (blue) at $U_{13}/t=87(10)$, $U_{12}/t=33(4)$, and $U_{23}/t=16(2)$, referring to interaction setting (III) of Fig.~\ref{fig: setup}{B}. The total density (black) is obtained by summing up all individual flavor densities. The experimental data (dots) is fitted to the high-temperature series expansions in the atomic limit (solid lines) using central chemical potentials and temperature as free parameters. The fit results are $\mu_{0,1}/t=17.2(7)$, $\mu_{0,2}/t=16.8(9)$, $\mu_{0,3}/t=17.1(9)$ and $k_B T/t =6.0(4)$. Error bars are one sigma fit errors. {\bfseries(B)} Doublon density as a function of total density for pairs, $\ket{1}-\ket{3}$ (violet), $\ket{1}-\ket{2}$ (orange), and $\ket{2}-\ket{3}$ (cyan). Dashed lines represent the theory with hopping included, which introduces excess doublons compared to the atomic limit \cite{SuppOnline}. {\bfseries(C)} On-site density variance. The suppression is observed at the density of one atom per lattice site, similar to the compressibility measurement in {\bfseries(D)}. Error bars are obtained by bootstrap analysis \cite{SuppOnline}.}
\label{fig: three-component MI}
\end{figure*}

To realize a three-component lattice gas, we prepare a balanced two-component Fermi gas using a $\ket{1}-\ket{3}$ mixture of $^6$Li atoms in a single layer of a one-dimensional accordion lattice. We obtain a balanced three-flavor mixture by applying radio-frequency (RF) pulses to drive the $\ket{1}-\ket{2}$ and $\ket{2}-\ket{3}$ transitions at the final stage of the evaporation \cite{SuppOnline}. We note that collisions between atoms within the timescale of the experiment result in an incoherent mixture of the flavors. Next, we tune the magnetic field to obtain the target interactions for the respective measurement. Atoms are loaded into a square lattice of a desired depth between 9.1(2)~$E_R$ and 16.0(3)~$E_R$ using an optimized s-shaped ramp \cite{SuppOnline}. Here, $E_R=\hbar^2\pi^2/(2ma^2_\text{latt})=h\times 14.6$ kHz is the recoil energy and $h$ is Planck's constant, $m$ is the atomic mass, and $a_\text{latt}=752$ nm is the lattice constant. The atom number and density in the lattice are adjustable by varying evaporation parameters. To image the atoms, the motion of the atoms is frozen by rapidly increasing the lattice to 50~$E_R$ in \SI{500}{\micro\second}. Next, we prepare for a Stern-Gerlach separation of two components by removing one flavor and mapping the remaining two flavors to a $\ket{2}-\ket{3}$ mixture. We freeze atomic motion in-plane by setting the in-plane lattice depth to 120~$E_R$ and turn on a vertical magnetic field gradient perpendicular to the lattice plane of approximately $170$ G/cm. Two oblate potentials (light sheets) are simultaneously turned on at a vertical distance of \SI{8}{\micro\meter}, capturing atoms in the states $\ket{2}$ and $\ket{3}$, respectively (Fig.~\ref{fig: setup}C). Both light sheets are further separated to \SI{20}{\micro\meter} for fluorescence imaging. This larger separation allows us to obtain two simultaneous site-resolved pictures of both planes on the same camera by splitting the fluorescence in two imaging paths and focusing each layer to a separate area of the camera \cite{SuppOnline}.


In a first measurement, we apply our flavor-resolved imaging to study the regime of strong repulsive interactions. For this measurement, we use a magnetic field of 625.9(3)~G corresponding to scattering lengths for the three pairs of flavors of $564(3)a_0$, $1509(16)a_0$, and $269(3)a_0$ for $a_{12}$, $a_{13}$ and $a_{23}$, respectively, where $a_0$ is the Bohr radius [(III) in Fig.~\ref{fig: setup}B]. We measure the flavor density, $n_\alpha$, for each flavor $\alpha$, as a function of radius from the trap center. The summation of these flavor densities gives the total density, $n=n_1+n_2+n_3$, as depicted in Fig.~\ref{fig: three-component MI}A. The density of sites with an atom of flavor $\alpha$ and an atom of flavor $\beta$ (doublon density), $n^d_{\alpha\beta}=\langle n_\alpha n_\beta \rangle$, can be reconstructed \cite{Koepsell2020}. Here, $\expval{\dots}$ denotes a quantum and thermal average. As can be seen in Fig.~\ref{fig: three-component MI}B, $n^d_{23}$ is larger than $n^d_{13}$ and $n^d_{12}$ because of lower interaction of $U_{23}$. We find generally good agreement with an atomic limit calculation \cite{SuppOnline}. We attribute the excess doublons of higher interaction pairs at high density to a small amount of hopping during the imaging \cite{SuppOnline}. A signature of a Mott insulator is the localization of atoms which can be detected through suppression of density variance and compressibility. Using our flavor-resolved data, we can extract these quantities (Fig.~\ref{fig: three-component MI}C-D), and we observed that the variance of the total density is suppressed at unit density (Fig.~\ref{fig: three-component MI}C) as expected for a Mott insulator \cite{SuppOnline}. Due to three-body losses at this magnetic field, accessing higher densities is challenging \cite{Ottenstein2008,Huckans2009,SuppOnline}. We confirmed the signature of the Mott insulator by studying the compressibility, $\kappa$, given by $\kappa n^2 = ({\partial n}/{\partial \mu})\rvert_T = -1/(m\omega^2 r) (\partial n/\partial r)$ \cite{SuppOnline}. In Fig.~\ref{fig: three-component MI}D, we observed reduced compressibility in the region of unit filling, suggesting insulating behavior in the regime of an ordinary unit-filling Mott insulator. The variance and compressibility do not completely vanish because of the finite temperature in our system which we attribute to heating caused by three-body collisions before loading into the lattice \cite{SuppOnline}. Overall the data is well described by a zeroth-order high-temperature series expansion (HTSE) with small deviations at the highest densities in the center of the trap \cite{SuppOnline}.

\begin{figure*}
\centering
\includegraphics[width=0.85\linewidth]{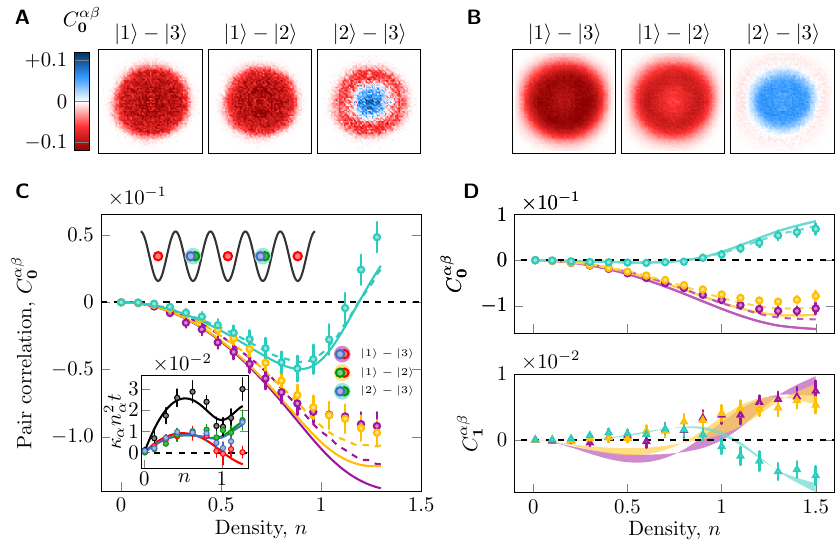}
\caption[Flavor-selective pairing]{{\bfseries Flavor-selective pairing.} {\bfseries(A,C)} Direct measurement of on-site pairing correlations in a Fermi lattice gas with interactions, $U_{13}/t=143(18)$, $U_{12}/t=53(7)$, and $U_{23}/t=26(3)$, referring to interaction setting (III) of Fig.~\ref{fig: setup}{B}. On-site pairing correlation maps are shown in {(A)}. Each map has a field of view of $80\times 80$ lattice sites. Azimuthal averaging of correlations {(C)} as a function of total density for $\ket{1}-\ket{3}$ (violet), $\ket{1}-\ket{2}$ (orange), and $\ket{2}-\ket{3}$ (cyan). Dots denote experimental on-site correlations. Solid lines represent the theoretical prediction from the zeroth-order HTSE fit to the flavor densities using central chemical potentials and temperature as free parameters. The fit results are $\mu_{0,1}/t=32(1)$, $\mu_{0,2}/t=30(3)$, $\mu_{0,3}/t=30(2)$, and $k_BT/t =10(1)$. Dashed lines include hopping in addition \cite{SuppOnline}. The top inset in {(C)} presents a likely atomic configuration in the lattice based on the measured correlations at $n > 1$ and the bottom inset illustrates the compressibility of each flavor (red, green, and blue) and the total density (black).
{\bfseries(B,D)} Measurement of the on-site and nearest-neighbor pairing correlations for interactions $U_{13}/t=13.7(5)$, $U_{12}/t=7.9(3)$, and $U_{23}/t=1.8(1)$ (interaction setting (II) of Fig.~\ref{fig: setup}{B}) with on-site pairing correlation maps {(B)}. The corresponding chemical potentials are $\mu_{0,1}/t=5.5(1)$, $\mu_{0,2}/t=4.2(1)$, $\mu_{0,3}/t=4.4(1)$, and the temperature is $k_BT/ t = 1.8(1)$. Azimuthal averaging of the on-site and nearest-neighbor (triangles) pairing correlations is shown in {(D)}. Shaded areas in (D) denote the theoretical expectations from NLCE using $6^\text{th}$ and $7^\text{th}$ order expansions as lower and upper limits and dashed lines in {(D, top)} take into account hopping in addition. Error bars are obtained from bootstrap analysis \cite{SuppOnline}. 
\label{fig: flavor-selective paring}}
  \end{figure*}
To study the correlations between two different flavors $\alpha$ and $\beta$, we define a pairing correlation given by
$C^{\alpha\beta}_{\mathbf{a}}(\mathbf{r})=\expval{n_{\alpha,\mathbf{r}}n_{\beta,\mathbf{r}+\mathbf{a}}}-\expval{n_{\alpha,\mathbf{r}}}\expval{n_{\beta,\mathbf{r}+\mathbf{a}}}$,
where $\mathbf{a}$ denotes the shift-vector between correlated positions, and $\mathbf{r}$ is the current lattice site \cite{Endres2013}. 
Therefore, for Fig.~\ref{fig: flavor-selective paring}A, we increased the center filling by $10\%$ by ramping the lattice $14\%$ deeper at the same magnetic field used in Fig.~\ref{fig: three-component MI}. This results in stronger confinement and higher central density while the atom number remains unchanged.
We observe suppression of $\ket{1}-\ket{3}$ and $\ket{1}-\ket{2}$ doublons for all densities but surprisingly, a crossover towards $\ket{2}-\ket{3}$ pairing in the center. For a more quantitative analysis, we perform azimuthal averaging (Fig.~\ref{fig: flavor-selective paring}C).
All on-site pairing correlations are negative for densities less than unity as expected from the previous observation of reduced compressibility in this regime. Surprisingly, we found a turning point of $C^{23}_0$ at unity filling and a crossover to positive correlation at $n\approx 1.1$, implying that $\ket{2}-\ket{3}$ pairs form at higher density while the remaining flavors still avoid each other. We find reduced correlations $C^{13}_0$ and $C^{12}_0$ compared to the theory at $n>1$, which we attribute to a small amount of hopping during the imaging \cite{SuppOnline}.
Additional evidence for flavor-selective pairing comes from a measurement of the flavor-density compressibility, $\kappa_\alpha n_\alpha^2 = ({\partial n_\alpha}/{\partial \mu})\rvert_T$ (Fig.~\ref{fig: flavor-selective paring}C,inset). In contrast to the compressibility of the flavors $\ket{2},\ket{3}$, the compressibility of flavor $\ket{1}$ for  $n > 1$ remains low instead of rising again, which is consistent with our theoretical model.

\begin{figure*}
\centering
\includegraphics[width=0.7\linewidth]{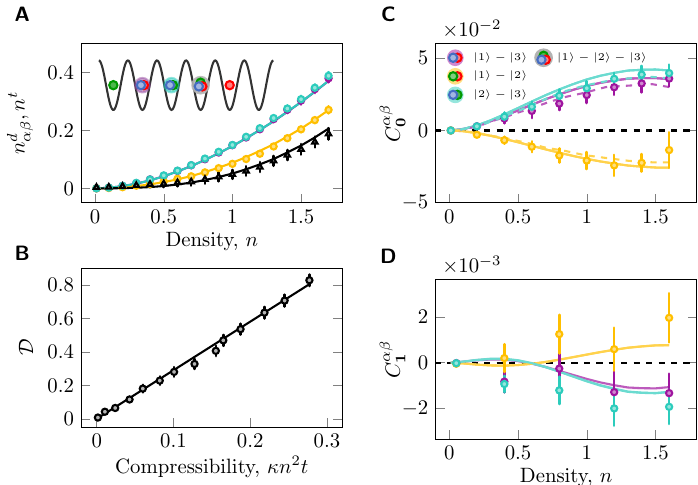}
\caption[Competing attractive pairing]{{\bfseries Competing attractive pairing.} On-site pair formation in a Fermi lattice gas with two comparable attractive and one repulsive interaction, $U_{13}/t=-2.4(1)$, $U_{12}/t=1.9(1)$, and $U_{23}/t=-2.7(1)$. These interactions correspond to the interaction setting (I) of Fig.~\ref{fig: setup}{B}. {\bfseries(A)} Doublon and triplon density. Circles repesent measured doublon densities and triangles measured triplon density. Lines are a NLCE fit with temperature and chemical potentials used as free variables. {\bfseries(B)} Total density fluctuation as a function of compressibility. The slope of the solid line represents the temperature of the system. Here, we obtain a temperature of $k_BT/t=2.9(2)$. {\bfseries(C,D)} Direct measurement of pairing correlations. The corresponding chemical potentials are $(\mu_{0,1},\mu_{0,2},\mu_{0,3})/t=(1.0(1),0.7(1),-0.4(1))$, and the temperature is $k_B T/ t = 3.0(1)$ from the NLCE fit of (A) using $6^\text{th}$ and $7^\text{th}$ order expansions (shaded solid lines). On-site pairing correlations {(C)} and nearest-pairing correlations {(D)}.
Hopping effects (dashed lines of the respective colors) are included in the NLCE in (C). Error bars are obtained by a bootstrap analysis \cite{SuppOnline}. The inset in {(A)} represents a likely atom configuration in the lattice which reproduces, when taking translational invariance into account, the signs of on-site and next-neighbor correlations.}
\label{fig: two attractive and one repulsive}
  \end{figure*}

To access higher densities and reach half-filling, we explore the regime where three-body loss is close to a minimum at approximately 570~G \cite{Ottenstein2008,Huckans2009,SuppOnline}. The magnetic field is ramped to the interaction setting (II) in Fig.~\ref{fig: setup}B and we set the lattice to 9.1(2) $E_R$, providing interaction strengths $U_{13}/t=13.7(5)$, $U_{12}/t=7.9(3)$, and $U_{23}/t=1.8(1)$. We immediately see from the spatially resolved correlation map (Fig.~\ref{fig: flavor-selective paring}B) that $\ket{2}-\ket{3}$ pairing occurs over almost the complete system size. In the azimuthal average, we see a positive on-site pairing correlation for $C^{23}_0$ for densities $n\gtrsim 0.8$ and negative on-site pairing correlations for $C^{13}_0$ and $C^{12}_0$ over the complete range of accessible densities (Fig.~\ref{fig: flavor-selective paring}D). The flavor-resolved quantum gas microscopy technique developed within this work allows to measure nearest-neighbor correlations between all flavors as well. All nearest-neighbor pairing correlations exhibit a turning point close to $n = 1$. 
To analyze these correlations, we fit the flavor densities to seventh-order site-expansion numerical linked-cluster expansion (NLCE) calculations, using temperature and chemical potentials at the trap center as free parameters \cite{SuppOnline}. The on-site and nearest-neighbor correlations are then calculated based on these fit parameters (Fig.~\ref{fig: flavor-selective paring}D).
Combining the results for on-site and nearest-neighbor pairing correlations suggests a tendency towards the ordering depicted in the insets of Fig.~\ref{fig: flavor-selective paring}C, where $\ket{2}-\ket{3}$ pairs and $\ket{1}$ atoms form a staggered ordering.
We attribute the  deviations between theory and experiment mainly to hopping during the imaging stage where the hopped atoms are most likely to form excess doublons  \cite{SuppOnline}.
Comparing both regimes represented in Fig.~\ref{fig: flavor-selective paring}A,C versus B,D, the relative interactions of different pairs are not significantly different but absolute interactions $U/t$ are reduced.

The tunability of interactions using the Feshbach resonances of lithium makes it even possible to study a three-flavor mixture with mixed-sign interactions.
We decided to study the competition between two types of pairing by exploring the interaction setting (I) in Fig.~\ref{fig: setup}B with two similar attractive and one repulsive interaction.  
Doublon densities are measured as a function of total density depicted in Fig.~\ref{fig: two attractive and one repulsive}A, showing competition between $\ket{1}-\ket{3}$ and $\ket{2}-\ket{3}$ pairs which have similar attractive interactions. As expected, we find higher densities for attractive pairs than for repulsive $\ket{1}-\ket{2}$ pairs.
By using the information gained from flavor-resolved imaging of all three pairs of flavors, we can even extract the density of triply occupied sites (triplon density), $n^t=\langle n_1 n_2 n_3 \rangle$, which is in good agreement with the theory. This demonstrates that these triply occupied sites do not lead to immediate atom loss. By measuring the on-site pairing correlations we find that attractive pairs are more likely to occupy the same site while repulsively interacting flavors avoid forming pairs, as expected (Fig.~\ref{fig: two attractive and one repulsive}C). The opposite signs of the on-site pair correlations are found in the nearest-neighbor pair correlations (Fig.~\ref{fig: two attractive and one repulsive}D). The repulsive pair prefers to occupy neighboring sites as opposed to attractive pairs which are found with reduced probability on neighboring sites. Combining the results from our correlation measurements, we find a tendency towards a state with a $\ket{1}-\ket{2}$ staggered ordering where $\ket{3}$ atoms are paired about equally likely to $\ket{1}$ or $\ket{2}$ atoms (inset of Fig.~\ref{fig: two attractive and one repulsive}A).

To independently cross-check our temperature measurements through theory fits, we define the total density fluctuation as
$
\mathcal{D} = \sum_{\mathbf{a}} (\expval{\hat{n}_\mathbf{r}\hat{n}_{\mathbf{r}+\mathbf{a}}}-\expval{\hat{n}_\mathbf{r}}\expval{\hat{n}_{\mathbf{r}+\mathbf{a}}}),
$
and extract the temperature independently relying on the density fluctuation-dissipation theorem \cite{Callen1951,Hartke2020} where temperature $T$ is given by $\kappa n^2 = \mathcal{D}/(k_B T)$. This allows us to confirm the consistency between both methods (Fig.~\ref{fig: two attractive and one repulsive}B).

\begin{figure}
\centering
\includegraphics[width=\columnwidth]{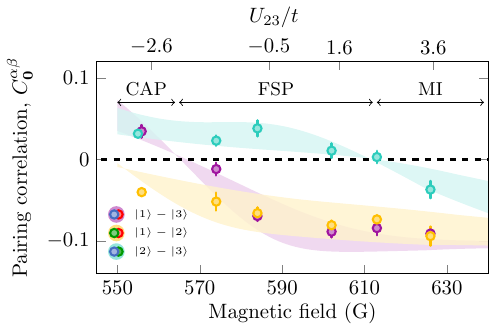}
\caption{{\bfseries Phase diagram of a balanced three-flavor Fermi Hubbard system.} Circles, magnetic field dependence of measured on-site pairing correlation at total density of $n=1.0(1)$ with equal population of all three flavors. The $\ket{1}-\ket{3}$ and $\ket{2}-\ket{3}$ pairing correlations show a crossover between pairing and anti-pairing over the accessible range. We find regimes of competing attractive pairing (CAP), flavor-selective pairing (FSP) and the Mott-insulating regime (MI). The shaded areas illustrate in-trap calculations of on-site correlations using a zeroth-order HTSE at unit center filling with temperatures between $k_BT/t =2$ and $k_BT/t =5$, and 850 atoms of each flavor \cite{SuppOnline}.}
\label{fig:phase diagram}
\end{figure}

Exploiting the complete range of tunability of interactions in our experiment, we summarize our findings for pairing in the three-flavor system for varying interactions in Fig.~\ref{fig:phase diagram}. We show the on-site pairing correlations for all three pairs at unity filling as a function of magnetic field at a fixed lattice depth of 9.1(2) $E_R$ \cite{SuppOnline}. $C^{23}_0$ exhibits a zero-crossing at a magnetic field of approximately 610 G, corresponding to $(U_{13},U_{12},U_{23})/t\simeq (13,7,2)$. This crossover represents the minimum interaction required to break all correlated pairs to reach the Mott regime at unit filling and sufficiently low tunneling.

In contrast, $C^{13}_0$ shows a zero-crossing at a magnetic field of approximately 568 G. This zero-crossing point corresponds to $(U_{13},U_{12},U_{23})/t\simeq (0,2.8,-1.7)$ and represents the maximum magnetic field at which the formation of $\ket{1}-\ket{3}$ pairs is preferred. 
We sort our data into three regimes using the sign of the on-site pairing correlations of the $\ket{1}-\ket{3}$ and $\ket{2}-\ket{3}$ pairs, $C^{13}_0$ and $C^{23}_0$: Mott insulating (MI), flavor-selective pairing (FSP), and competing attractive-pairing (CAP).

For $C^{13}_0 < 0$ and $C^{23}_0 < 0$ we find the MI state [Fig.~\ref{fig: flavor-selective paring}(A,C)], where all on-site pairs are suppressed. For $C^{13}_0 < 0$ and $C^{23}_0 > 0$ we find the FSP state which shows a significant amount of $\ket{2}-\ket{3}$ doublons at sufficiently high densities [Fig.~\ref{fig: flavor-selective paring}(B,D)].
In the CAP regime there is a competition in the formation of $\ket{1}-\ket{3}$ doublons and $\ket{2}-\ket{3}$ doublons and we find a significant triplon fraction (Fig.~\ref{fig: two attractive and one repulsive}).

In conclusion, we demonstrated flavor-resolved quantum gas microscopy of three-flavor Fermi lattice gases in the Hubbard regime. We observe the onset of three-flavor Mott insulators, flavor-selective localization, and selective pairing at temperatures down to the tunneling scale through the direct detection of flavor densities, pairing correlations and triply occupied sites. We find overall good agreement with NLCE calculations and a zeroth-order HTSE with small deviations at high densities. 

Limitations preventing us from achieving three-flavor Mott insulators at half-filling with temperatures lower than the tunneling scale are attributed to three-body losses at a magnetic field of approximately $626$ G. To suppress three-body losses, it may help to significantly increase the confinement of the vertical 1d lattice, referred to as the accordion lattice above, to boost the interaction strength while staying in the minimum three-body loss regime around $600$ G. Overcoming these limitations may enable a further increase in phase-space density and the observation of a novel paired Mott phase at half-filling.

Our system opens up the ability to study three-flavor fermionic systems in other optical lattice geometries. In particular, triangular and kagome lattices may show exotic phases such as chiral states that break time-reversal symmetry \cite{Boos2020,Hafez-Torbati2020,Xu2023_SU3}. In addition, a three-flavor Fermi systems with ultracold atoms provide a pathway for the study of color superfluidity and aspects of quantum chromodynamics \cite{Rapp2007,Banerjee2013,Tajima2022, Xu2023_trion}.

%

\subsection*{Acknowledgements}

We thank Richard Scalettar for discussions.
We thank Waseem Bakr and his group for discussions on state-resolved imaging techniques.
P.S., J.M. and L.L. acknowledge support by the National Science Foundation (CAREER award \#2047275), the Thomas F. and Kate Miller Jeffress Memorial Trust and the Jefferson Trust. J. M. acknowledges support by The Beitchman Award for Innovative Graduate Student Research in Physics in honor of Robert V. Coleman and Bascom S. Deaver, Jr.
K.R.A.H. and S.D. acknowledge support from the National Science Foundation (PHY-1848304 and QIS-2346014), the Robert A. Welch
Foundation (C-1872), and the W. M. Keck Foundation (Grant No. 995764).

\clearpage

\section*{Appendix}

\section{EXPERIMENTAL METHODS}
\subsection{Preparation of three-flavor Fermi gases}
\begin{figure}[ht]
    \centering
    \includegraphics[width=0.9\linewidth]{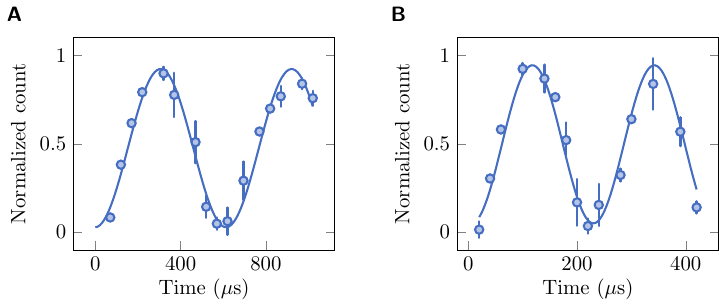}
    \caption{{\bfseries Rabi oscillations of three lowest hyperfine states.} {(\textbf{A})} $\ket{1}-\ket{2}$ oscillation with a Rabi frequency of $(2\pi)\times 1.63(8)$ kHz and {(\textbf{B})} $\ket{2}-\ket{3}$ state with a Rabi frequency of $(2\pi)\times4.46(21)$ at a magnetic field of 594~G. Error bars represent the standard error of the mean over three datasets.}
    \label{fig:Rabi oscillation}
\end{figure}

To prepare a three-component mixture, we ramp the magnetic field to 594~G in 50~ms at the end of the final evaporation and obtain a tunable mixture by applying radio-frequency (RF) pulses to drive the $\ket{1}-\ket{2}$ and $\ket{2}-\ket{3}$ transitions. The optimization of a balanced three-component mixture is accomplished by fine-tuning the pulse duration to transfer one third of the population in each pulse (Fig.~\ref{fig:Rabi oscillation}).

\begin{figure}[ht]
    \centering
    \includegraphics[width=0.9\columnwidth]{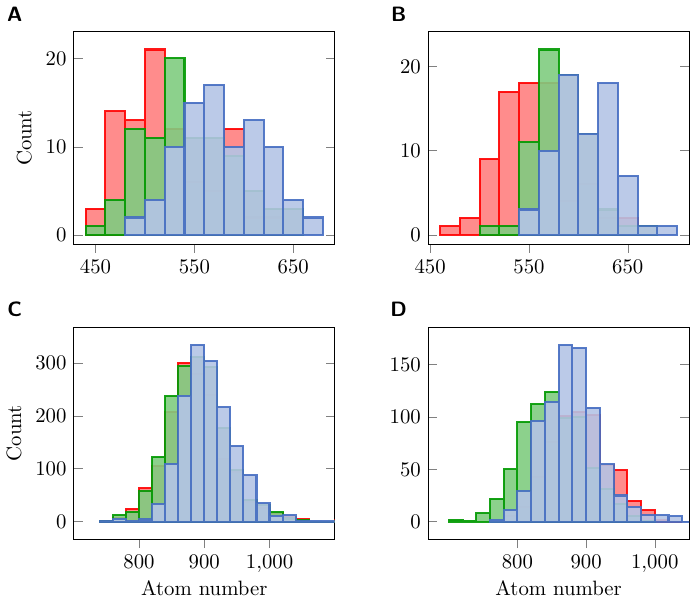}
    \caption{{\bfseries Atom number histograms.} {(\textbf{A} to \textbf{D})} correspond to the Figs.~2, 3A, 3B, and 4 of the main text, with atom numbers per flavor as follows: $(524\pm 45,543\pm 45,582\pm 45)$,$(554\pm 35,581\pm 27,608\pm 31)$,$(889\pm 52,889\pm 45,907\pm 43)$, and $(891\pm 42,852\pm 45,879\pm42)$.}
    \label{fig: atom number histogram}
\end{figure}

We verified the flavor population for each dataset by constructing atom number histograms as demonstrated in Fig.~\ref{fig: atom number histogram}. 

\begin{figure}
    \centering
    \includegraphics[width=0.6\columnwidth]{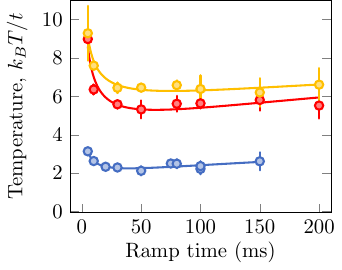}
    \caption{{\bfseries Fit temperature as a function of lattice ramp time.} Dots represent data for lattice depths of \SI{9.1(2)}{\Ersq} (blue), \SI{12.0(2)}{\Ersq} (red), and \SI{14.0(3)}{\Ersq} (orange). Solid lines are fits to polynomial functions. The ramp time used for the measurements in the main text is chosen based on the minimized fit temperature. Error bars are calculated using bootstrap analysis from 60 datasets.}
    \label{fig:T vs ramp time}
\end{figure}

The thermalization dynamics of three-flavor Fermi gases has not been studied in detail previously. To make sure the thermalization before loading into the lattice does not impact our measurements significantly, we vary the duration of the loading into a square lattice and measure all three flavor densities. The density profiles are fitted to zeroth-order high temperature series expansions to extract an effective temperature (Sec.~\ref{section HTSE0}). The experimental ramp time is determined based on the minimal fitted effective temperature, as depicted in Fig.~\ref{fig:T vs ramp time}. We note that shorter ramp durations lead to non-equilibrium states which the theory fails to capture, while longer ramp durations suppress the correlations because of higher temperatures caused by heating of the Fermi gas. Table~\ref{table: parameters used in experiment} shows the experimental parameters used in the main text.

\begin{table}
    \centering
    \begin{tabular}{||c c c c||} 
     \hline
     Figure & Depth (\SI{}{\Ersq}) & Field (G) & $(U_{13},U_{12},U_{23})/t$ \\ [0.5ex] 
     \hline\hline
     2 & 14.0 & 625.9 & $(87,33,16)$ \\
     \hline
     3A & 16.0 & 625.9 & $(143,53,26)$ \\
     \hline
     3B & 9.1 & 603.3 & $(13.7,7.9,1.8)$ \\
     \hline
     4 and 5 & 9.1 & 555.9 & $(-2.4,1.9,-2.7)$ \\ 
    \hline
      5 & 9.1 & 574.2 & $(1.6, 4.3, -1.8)$ \\
       \hline
      5 & 9.1 & 584.1 & $(4.6, 4.8, -0.5)$ \\
       \hline
      5 & 9.1 & 603.3 & $(13.7,7.9,1.8)$ \\
       \hline
      5 & 9.1 & 613.0 & $(18.3,8.6,3.1)$ \\
     \hline
      5 & 9.1 & 625.9 & $(21.8, 8.1, 3.9)$ \\ 
     \hline
    \end{tabular}
 \caption{{\bfseries Summary of parameters used in the main text.}}
 \label{table: parameters used in experiment}
\end{table}

\subsection{Bilayer implementation and readout}
\begin{figure}[ht]
    \centering
    \includegraphics[width=\linewidth]{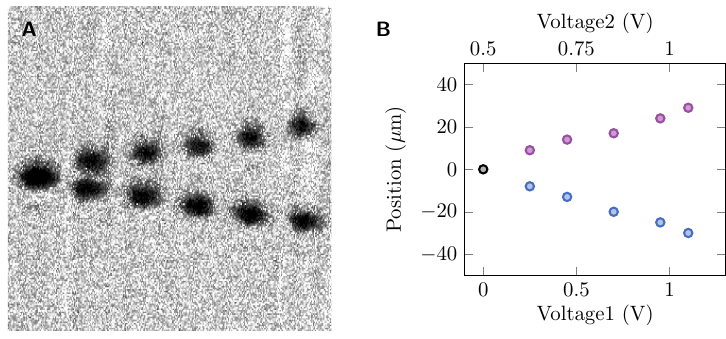}
    \caption{{\bfseries Bilayer imaging using two light sheets.} {(\textbf{A})} Composition of absorption images of the light sheet from the side. The two light sheets are initially overlapped at the same position (left, 0 V) and are then split vertically by apply increasing voltages simultaneously to both galvanometers. By varying the galvanometer voltages, both potentials can be individually controlled and here we move them away from each other. The field of view is \SI{200}{\micro\meter} $\times$ \SI{200}{\micro\meter}. The separation between both light sheets can be greater than \SI{50}{\micro\meter}. {(\textbf{B})} Light sheet positions as a function of voltage applied to the galvanometers. Blue (violet) dots are potentials that capture states $\ket{2} (\ket{3})$. The black dot represents the parameters at which the two light sheets are located at the same position.}
	\label{fig: double light sheets setup}
    \label{fig:enter-label}
\end{figure}

The setup of the two light sheets is shown in Fig.~1 of the main text. We split the power 50:50 between the two light sheets using a high-power polarizing beamsplitter cube and combine both light sheets with crossed polarization on another beamsplitter. We add a 1:1 telescope to the longer path, thus ensuring the beams are of identical shape. Atoms are captured by the two light sheets after a Stern-Gerlach separation. During the imaging process, we collect scattered light from atoms in both light sheets simultaneously through the high-resolution objective and split the fluorescence light of the two individual light sheet layers equally by a beamsplitter. We image both layers at the same time by compensating the focus shift of the reflected beam from the beamsplitter by a thick glass block. The vertical alignment of both light sheets can be individually computer-controlled by two mirrors attached to galvanometer scanners (Fig.~\ref{fig: double light sheets setup}).

\begin{figure}[ht]
	\centering
    \includegraphics[width=\linewidth]{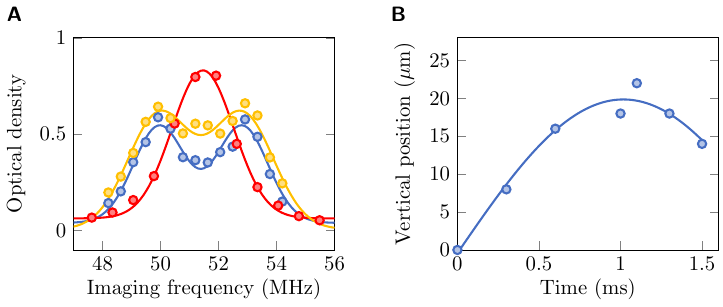}
    \caption[Zero field measurement and oscillation of state $\ket{3}$]{{\bfseries Zero field measurement and oscillation of state $\ket{3}$.} {(\textbf{A})} Optical density as a function of low field imaging frequency in the presence of various magnetic fields. Background without applied fields (red), anti-Helmholtz of Feshbach field at \SI{92}{\ampere} (blue), Z-offset field at \SI{10}{\ampere} (orange). 
    The peak separation indicates the Zeeman splitting, which allows us to infer the magnetic offset field. The Zeeman splitting is minimized when the magnetic field at the atom position is zero. We find a residual offset field at the atom position of 5 G when applying only the gradient. This offset field of 5~G is maintained during the splitting process to avoid spin flips. Therefore, we do not need additional offset fields during the splitting process.
    {(\textbf{B})} Classical dynamics. The cloud is initially held in a lattice with a depth of \SI{16}{\Ersq}. By applying a magnetic field gradient of \SI{35}{G/cm}, oscillations occur around a new equilibrium determined by the vertical harmonic confinement of the lattice and the gradient force.
    By fitting the oscillation to a sinusoidal function, the oscillation frequency is extracted and yields approximately $(2\pi)\times$\SI{300(29)}{\hertz} for the vertical trap frequency in the lattice tubes. This result is consistent with the measurement of non-interacting Fermi gas in Sec.~\ref{Confinement calibration}.}
\label{fig: Zero field and spin3 oscillation}
\end{figure}

To perform the Stern-Gerlach separation, the control of the magnetic field gradient is key. We form an anti-Helmholtz coil configuration by reversing the current direction through one Feshbach coil using a water-cooled H-bridge circuit. This allows us to access magnetic fields gradients of up to \SI{280}{G/cm} at \SI{200}{\ampere}. 
The states we choose to perform the separation are $\ket{2}$ and $\ket{3}$ because the state $\ket{3}$ has a strong magnetic moment at almost zero field and the sign is opposite to the magnetic moment of $\ket{2}$. Therefore, both components will split efficiently in the presence of a magnetic field gradient. 
We ensure that we maintain a small magnetic offset field of about $5$~G during the splitting to maximize the magnetic moment difference between both states (Fig.~\ref{fig: Zero field and spin3 oscillation}A).
We verified the direction of separation by initially holding a $\ket{2}-\ket{3}$ mixture in an optical lattice and then applying a constant magnetic field gradient. By observing the oscillation of the atomic cloud in the $\ket{3}$ state around a new equilibrium under the influence of the lattice and gradient, we see that the vertical oscillations follow a simple harmonic motion around the trap center, as shown in Fig.~\ref{fig: Zero field and spin3 oscillation}B.

Independent of the atomic states used for the Hubbard physics, we always map a pair of flavors for imaging on a $\ket{2}-\ket{3}$ mixture. To obtain this $\ket{2}-\ket{3}$ mixture that is suitable for the Stern-Gerlach separation, we ramp the lattice to \SI{50}{\Ersq} and apply a $\ket{1}-\ket{2}$ RF sweep centered at \SI{594}{G} spanned by \SI{50}{\kilo\hertz}. This transfer has an efficiency greater than 99\%. Next, we ramp up the lattice depth to \SI{120}{\Ersq} and turn on the magnetic field gradient of approximately \SI{170}{G/cm}. This magnetic field gradient is chosen to provide a convenient steady-state splitting of the two spin components in the vertical direction in the harmonic confinement of the optical lattice at \SI{120}{\Ersq}. Two light sheets are simultaneously turned on and stay \SI{8}{\micro\meter} apart from each other, capturing atoms occupying in states $\ket{2}$ and $\ket{3}$, respectively. We then move both light sheets further apart to \SI{20}{\micro\meter} and perform fluorescence imaging as described in ref.~\cite{Yang2021}.

\begin{figure}
    \centering
    \includegraphics[width=\linewidth]{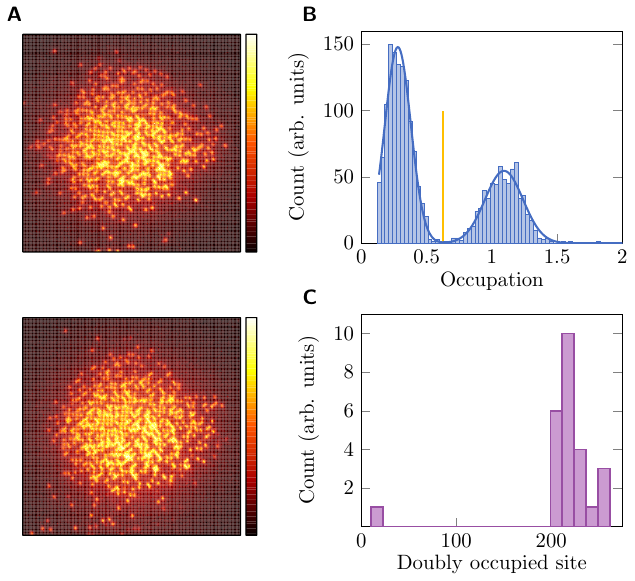}
    \caption{{\bfseries Flavor-resolved imaging.} {(\textbf{A})} Single-site resolution of $\ket{1}$ (top) and $\ket{3}$ (bottom). {(\textbf{B})} Count histogram of reconstruction in flavor-resolved imaging. Orange marks the optimized threshold of occupation. {(\textbf{C})} Histogram of double occupations for various shift vectors. By overlapping the center of $\ket{1}$ and $\ket{3}$, a certain configuration shows an extremal number of doublons which is the correct overlapping shift vector.}
    \label{fig:Spin resolved imaging}
\end{figure}

In Fig.~\ref{fig:Spin resolved imaging}A, we show flavor-resolved fluorescence images of atoms in a two-component system that show an antiferromagnetic patch as expected for a square lattice in a repulsive regime. We suppress the background noise from the other layer by increasing the exposure time to \SI{2}{\second} and the resulting count histogram for the atom signal per lattice site shows well-separated peaks between empty and occupied sites (Fig.~\ref{fig:Spin resolved imaging}B). The center position of both flavor-resolved pictures needs to be recovered by an overlapping procedure. For example, the accuracy of overlapping using the center of mass has an error of a few sites, typically less than two lattice sites. To overlap more precisely, we apply small shifts within the error margin of the center-of-mass overlapping procedure. We then select the shift vector that results in an extremal doublon number, distinct from the noise associated with other shift vectors, similar to the approach in ref.~\cite{Koepsell2020}. We only use pictures in the evaluation where this procedure provides a unique result.

\subsection{Imaging fidelity}\label{sec:imaging_fidelity}
\begin{figure}[ht]
    \centering
    \includegraphics[width=0.6\linewidth]{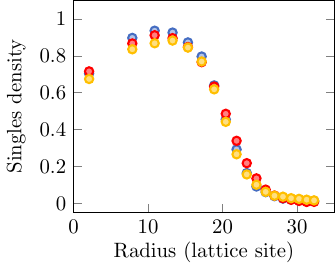}
    \caption{{\bfseries Stern-Gerlach separation fidelity.} Singles densities without and with splitting are $93.6(3)\%$ (blue), $91.1(4)\%$ (red). When reversing the splitting process, a lower singles density of $88.2(4)\%$ is obtained (orange).}
    \label{fig:fidelity of splitting}
\end{figure}

We measure the splitting fidelity by preparing a two-component Mott insulator with a maximum probability of singly occupied sites in a radial bin of 93.6(3)\% and compare it to a flavor-resolved image with a filling of 91.1(4)\% (Fig. \ref{fig:fidelity of splitting}). The reduction in density of singly occupied sites indicates losses during the transport process. 

Imaging fidelity without and with splitting are $95(1)\%$, $94(1)\%$ by taking a series of images similar to previous work \cite{Yang2021}. The imaging fidelity is slightly lower than in our previous work because we increased the exposure time to compensate for the background noise from the defocused layer.

\section{CALIBRATION OF HUBBARD PARAMETERS}
\subsection{Tunneling calibration}
\begin{figure}[ht]
    \centering
    \includegraphics[width=0.9\linewidth]{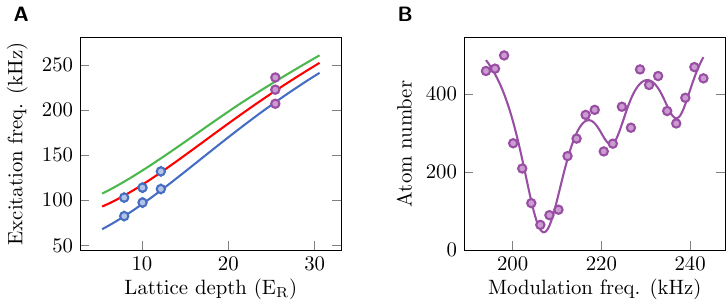}
    \caption{{\bfseries Lattice Calibration.} {(\textbf{A})} Band excitation resonances as a function of lattice depth. Dots are measurements and solid lines are band transitions calculated from the band structure in the tight-binding limit. Third resonances are too weak for reliable identification for shallow lattice depths. {(\textbf{B})} Lattice amplitude modulation data corresponding to the violet dots in {({A})}. Data (dots) is fit to three Gaussian functions (solid line). With the knowledge of band transition frequencies, the experimental lattice depths are determined.}
    \label{fig:lattice calibration}
\end{figure}

We prepare non-interacting Fermi gases of a $\ket{1}-\ket{3}$ mixture in a square lattice and perform amplitude modulation for \SI{100}{\milli\second} with an amplitude modulation of 5\% for varying modulation frequencies. We observed atom losses near the transition of the s-band to the d-band (Fig.~\ref{fig:lattice calibration}). This data allows us to extract lattice depth for a certain lattice beam power in analogy to previous work \cite{Mongkolkiattichai2023}.

\subsection{Confinement calibration \label{Confinement calibration}}

To determine our radial confinement in the lattice we fit the density profile of a non-interacting Fermi gas. In this non-interacting system, we lose the interaction as a fit parameter and can use the confinement as a free parameter instead. In the calculation, we fill up the Fermi-Dirac distribution with the energy calculated from the tight-binding model and take into account a local density variation due to the confinement, $\mu(r)$. For the square lattice, the energy dispersion is given by
\begin{equation}
\varepsilon^\text{sq}(\mathbf{k})=-2t[\cos(k_xa)+\cos(k_ya)]
\end{equation}
where $t$ is the tunneling strength, $k_i$ is the lattice momentum defined in the first Brillouin zone, $k\in [-\pi/a,\pi/a)$. $a$ is the lattice spacing.

The density profile of a non-interacting gas in the square lattice can be calculated by summing up all allowed momenta
\begin{equation}
n^\text{sq}_\text{non-int}(r)=\frac{1}{(2\pi)^2}\int_{-\pi/a}^{\pi/a}\int_{-\pi/a}^{\pi/a}\frac{1}{e^{\beta(\varepsilon^\text{sq}(\mathbf{k})-\mu(r))}+1}\ dk_x dk_y
\label{eq: noninteracting fermi gas}
\end{equation}

\begin{figure}[ht]
    \centering
    \includegraphics[width=0.6\linewidth]{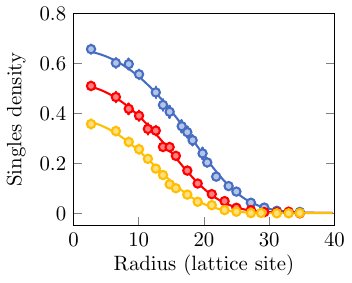}
    \caption{{\bfseries Azimuthal average of non-interacting gas.} The density profiles are fit to Eq.~\ref{eq: noninteracting fermi gas} and we obtain $k_BT/t\sim 1$ with the central chemical potentials of $\mu_0/t$ of $1.2$ (blue), $0.2$ (red), $-0.7$ (orange), leading to confinement of $(2\pi)\times$\SI{166(7)}{\hertz} for the lattice depth of \SI{7.4}{\Ersq}.}
    \label{fig: confinement non interacting gas}
\end{figure}

We obtain a non-interacting Fermi gas similar to the Mott insulator sequence using a $\ket{1}-\ket{2}$ mixture. In addition, we ramp up the Feshbach field to the non-interacting point at \SI{527}{G} after evaporation in the accordion lattice with horizontal confinement provided by a Gaussian beam propagating upward, referred to as the bottom beam. Then the lattice depth is set to \SI{7.4}{\Ersq}. The azimuthal average of the cloud is shown in Fig.~\ref{fig: confinement non interacting gas}.

\subsection{Interaction calibration}
\begin{figure}
    \centering
    \includegraphics[width=0.9\linewidth]{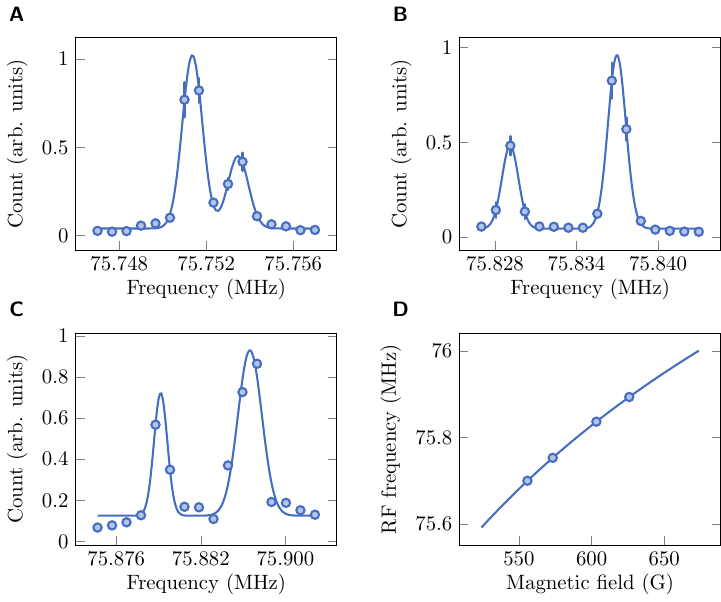}
     \caption{{\bfseries Interaction Calibration.} {(\textbf{A} to \textbf{C})} Radio-frequency spectra at magnetic offset fields of \SI{573.3(3)}{G}, \SI{603.3(1)}{G}, and \SI{625.9(3)}{G} using lattice depth of \SI{15.0(3)}{\Ersq}. The right peak corresponds to the transfer of $\ket{1}$ from singly occupied sites and the left peak is due to the transfer of $\ket{1}$ in doubly occupied sites. We can infer the magnetic field using the right peak and the peak separation is proportional to interaction in analogy to the calibration in ref. \cite{Mitra2018}.
     The peak separations in {({A} to {C})} are $2.10(7)$, $7.85(8)$, and \SI{14.8(16)}{\kilo\hertz}. (\textbf{D}) Energy separation for the $\ket{1}$-$\ket{2}$ transition as a function of magnetic field. Dots are measurements obtained from {({A} to {C})}. The leftmost point is for a field of \SI{555.9(1)}{G}. Solid line is calculated using the Breit-Rabi formula. 
     }
    \label{fig:U calibration}
\end{figure}

To obtain interactions between flavor pairs, we prepare a $\ket{1}-\ket{3}$ band insulator at a target field using a lattice depth of \SI{15}{\Ersq}. We transfer atoms from state $\ket{1}$ to $\ket{2}$ using an RF-pulse with a \SI{1}{\milli\second} duration. By scanning the RF frequency, the $\ket{1}$ atoms in singles and doublons are transferred to ${\ket{2}}$ at different frequencies (Fig.~\ref{fig:U calibration}). The separation of the two peaks allows us to determine the interaction of the $\ket{1}-\ket{3}$ pair. The interactions of the remaining pairs are inferred from the field-dependent scattering lengths reported in ref.~\cite{Zurn2013}.

\subsection{Field-dependence of three-flavor Fermi gas loss rates}
\begin{figure}[ht]
    \centering
    \includegraphics[width=0.6\linewidth]{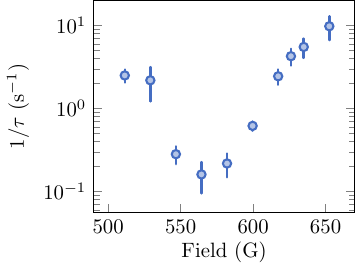}
    \caption{{\bfseries Loss rate of state $\ket{1}$ in a three-component mixture.} Each data point is obtained by fitting the atom number versus hold time to an exponential decay. Error bars are extracted from a nonlinear least squares fitting. Similar results are observed for states $\ket{2}$ and $\ket{3}$ (not shown).}
    \label{fig:123 lifetime}
\end{figure}

The magnetic-field-dependent stability of three-flavor Fermi gases is investigated by holding the gases in the accordion lattice confined horizontally by the bottom beam at the end of the final evaporation for varying duration. The state $\ket{1}$ is observed through fluorescence imaging in the lattice. To determine the lifetime, we fit an exponential decay to the measurements and obtain the lifetime as a function of magnetic field, as depicted in Fig.~\ref{fig:123 lifetime}.

\section{HIGH-TEMPERATURE SERIES EXPANSION \label{section HTSE0}}
In the regime where interactions greatly exceed tunneling ($U\gg t$), the singles occupation, $n^s$, can be calculated by the grand canonical ensemble, given by
\begin{equation}
    n^s = \frac{1}{Z}\sum_{n_1= 0}^{1}\sum_{n_2 = 0}^{1}\sum_{n_3= 0}^{1} \mod({n_1+n_2+n_3,2}) \ z(n_1,n_2,n_3)
\end{equation}
where $Z$ is the partition function:
\begin{equation}
    Z = \sum_{n_1= 0}^{1}\sum_{n_2 = 0}^{1}\sum_{n_3= 0}^{1} z(n_1,n_2,n_3),
\end{equation}
and
\begin{equation}
    z(n_1,n_2,n_3)=\exp(\beta\left(\sum_{\alpha=1}^{3} \mu_\alpha n_\alpha - \frac{1}{2} \sum_{\alpha\neq\gamma} U_{\alpha\gamma}n_\alpha n_\gamma\right)).
\end{equation}
Here $\beta = 1/(k_B T)$ and $k_B$ is the Boltzmann constant. The function $\mod{(\dots)}$ is the modulo operation that projects the total density to 0 or 1. The physical reason are light-assisted collisions, causing the loss of doubly occupied sites during imaging.

\noindent Single-component densities of three-component mixture, $n_i$, are simultaneously calculated by
\begin{equation}
    [\bar{n}_1,\bar{n}_2,\bar{n}_3] = \frac{1}{Z}\sum_{n_1= 0}^{1}\sum_{n_2 = 0}^{1}\sum_{n_3= 0}^{1}
    [n_1,n_2,n_3]\ z(n_1,n_2,n_3),
\end{equation}
and mean squared of single-component densities are given by
\begin{equation}
    [\bar{n^2_1},\bar{n^2_2},\bar{n^2_3}] = \frac{1}{Z}\sum_{n_1= 0}^{1}\sum_{n_2 = 0}^{1}\sum_{n_3= 0}^{1}
    [n_1^2,n_2^2,n_3^2]\ z(n_1,n_2,n_3).
\end{equation}
The variance of the total density is calculated by
\begin{align}
    \sigma^2 &= \bar{n^2} - \bar{n}^2 =\sum_{\alpha=1} \bar{n^2_\alpha} - \left(\sum_{\alpha=1} \bar{n}_\alpha\right)^2.
\end{align}
Doublon densities of the three pairs are calculated by
\begin{equation}
    \begin{aligned}
        &[\bar{n}^d_{13},\bar{n}^d_{12},\bar{n}^d_{23}] = \\
        &\frac{1}{Z}\sum_{n_1= 0}^{1}\sum_{n_2 = 0}^{1}\sum_{n_3= 0}^{1}
    [n_1n_3,n_1n_2,n_2n_3] z(n_1,n_2,n_3).
    \end{aligned}
    \label{eq: doublon density from HTSE}
\end{equation}
The triplon density is calculated by
\begin{equation}
    n^t = \frac{1}{Z}\sum_{n_1= 0}^{1}\sum_{n_2 = 0}^{1}\sum_{n_3= 0}^{1}n_1n_2n_3 z(n_1,n_2,n_3).
\end{equation}
According to Eq.~\ref{eq: doublon density from HTSE}, we define the doublon density as the number pairs of particles of a certain kind that are contained in a lattice site. This definition includes lattice sites where there are exactly two particles (doublons), as well as lattice sites with three particles (triplons). Consequently, sites with three particles are counted as having a doublon in this density measure, which means that the density includes multiple counts of such sites. We note that only the dataset shown in Fig.~4 of the main text shows a significant number of triplons.

\begin{table*}
    \centering
    \begin{tabular}{|c|c|}
    \hline
         $(U_{13}, U_{12}, U_{23})/t$ & $(-2.4(1), 1.9(1), -2.7(1))$\\
         \hline \hline
         Atomic Limit & $T/k_Bt=3.23(2)$, ${\vec\mu}/t=(0.98(4),0.69(4),-0.39(4))$ \\
         \hline         
         NLCE ($7^\text{th}$ order)& $T/k_Bt=2.98(2)$, ${\vec\mu}/t=(0.96(4), 0.68(4), -0.39(4))$ \\
         \hline 
          &  \\
         \hline
         $(U_{13}, U_{12}, U_{23})/t$ & $(13.7(5), 7.9(3), 1.8(1))$\\
         \hline \hline
         Atomic Limit & $T/k_Bt=2.10(7)$, ${\vec\mu}/t=(5.4(1), 4.2(1), 4.4(1))$ \\
         \hline         
         NLCE ($7^\text{th}$ order)&  $T/k_Bt=1.8(1)$, ${\vec\mu}/t=(5.5(1), 4.2(1), 4.4(1))$\\
         \hline
    \end{tabular}
    
    \caption{Summary of the fit parameter values for all NLCE calculations. 
    }
    \label{table: fit parameters for NLCE}
\end{table*}

\section{NUMERICAL LINKED-CLUSTER EXPANSIONS}
A seventh-order site-expansion numerical linked-cluster expansion (NLCE) is used to compute the observables in the main text.
With the NLCE algorithm \cite{Rigol2006,Tang2013}, a thermodynamic property $P(\mathcal{L})$ of the lattice $\mathcal{L}$ is written as a weighted sum of the property value for all its connected clusters,
\begin{equation}\label{eq:nlce-sum}
    P(\mathcal{L})/N_s = \sum_{c\subset \mathcal{L}} L(c)  W_P(c),
\end{equation}
where $L(c)$ is the number of ways cluster $c$ can be embedded in $\mathcal{L}$ up to translation invariance and topological equivalence, and the property weight $W_P(c)$ of cluster $c$ is defined recursively as~\footnote{Note that Eq.~\ref{eq:nlce-sum} follows from setting $c=\mathcal{L}$.}
\begin{equation}
    W_P(c) = P(c) - \sum_{s\subset c} W_P(s)
\end{equation}
where the sum is over all subclusters $s$ of $c$.
$P(c)$ is computed by exact diagonalization (ED).
The site-expansion scheme \cite{Tang2013} groups terms with an equal number of sites in a cluster (the order of the expansion) in the sum of Eq.~\eqref{eq:nlce-sum}.
When the correlations are short-ranged, the sum can be truncated to a reasonably low order with errors exponentially small in the truncation order.

The only error the NLCE suffers is the truncation error, which is controlled by the truncation order and is self-diagnosed by comparing consecutive orders.
Results are significantly more accurate than bare ED of equivalent system sizes. 
See Appendix~B of Ref.~\cite{Ibarra2021} for a thorough comparison in the closely related systems of  SU($N$) Fermi-Hubbard models. We use the $U(1) \times U(1) \times U(1)$ symmetry (i.e. conservation of each spin species), as in the SU(3) calculations, but the absence of SU(3) symmetry in the present system prevents the use of permutation symmetry, requiring diagonalization of up to six times as many clusters. In Table~\ref{table: fit parameters for NLCE} we summarize the fit parameters used in Figs.~3D and 4 of the main text.

\section{HOPPING CORRECTION}
For Figs.~2B,~3B,~3D, and 4 of the main text, we attribute the deviation from the theory calculations predominantly to hopping during the imaging stage in the experiment.
We calibrated independently our imaging fidelity of the flavor-resolved imaging to \SI{94}{\percent} (Sec.~\ref{sec:imaging_fidelity}). The infidelity is mostly caused by hopping between different vertical tubes of the lattices during the splitting, and to a much smaller extent by losses. We find for the hopping in this flavor-resolved imaging process \SI{5(1)}{\percent}.
To take the effect of this hopping during the imaging process into account to the first order in the numerical data, we use a Monte Carlo technique as follows: We generate $5000$ samples of $20\times20$ atom distributions which are generated with spatially homogeneous probabilities for all eight possible on-site occupations (empty,$\ket{1}$,$\ket{2}$,$\ket{3}$,$\ket{1}\ket{3}$,$\ket{1}\ket{2}$,$\ket{2}\ket{3}$,$\ket{1}\ket{2}\ket{3}$). We then iterate over all atoms and introduce a hop event with a $p = 5 \%$ probability which is distributed equally to the four nearest-neighbor sites (each with $p/4$ probability). The pairing observables are thereafter calculated by averaging from the resulting $5000$ samples with hopping using the formulas discussed in Sec.~\ref{section HTSE0}. We observe that the number of samples is sufficient to reproduce the data without hopping within the size of the symbols in the figures of the main text. We note that this hopping correction is only a first-order approximation and does not take into account many other possibly relevant effects, including nearest-neighbor correlations, interactions, and density-dependent hopping.

\section{DATA ANALYSIS}

\subsection{Total three-flavor density variance}
We evaluate the total density variance, $\sigma^2$, from the experiment by expanding
\begin{equation}
     \sigma^2 = \expval{(n_1+n_2+n_3)^2}-\expval{n_1+n_2+n_3}^2.
\end{equation}
By defining
\begin{equation}
    \sigma^2_{\expval{n_in_j}} = \expval{n_i n_j} - \expval{n_i}\expval{n_j} = C^{ij}_0,
\end{equation}
the variance can be written in terms of observable quantities

\begin{equation}
    \sigma^2 = C^{11}_0+C^{22}_0+C^{33}_0+2(C^{13}_0+C^{12}_0+C^{23}_0).
    \label{eq:variance formula}
\end{equation}
We note that all terms in Eq.~\ref{eq:variance formula} are accessible through our flavor-resolved imaging.

\subsection{Compressibility}
By applying local density approximation, the chemical potential can be written in terms of radial dependence. The compressibility, $\kappa$, is given by
\begin{equation}
    \kappa n^2 = \left.\frac{\partial n}{\partial \mu}\right\rvert_T = -\frac{1}{m\omega^2 r} \frac{\partial n}{\partial r}
\end{equation}
where $n$ is the total density extracted from individual measurements of the three
flavors and $\omega$ is the lattice confinement obtained from the measurement in Sec.~\ref{Confinement calibration} \cite{Ku2012}.

Similarly, we calculate the compressibility $\kappa_\alpha$ of the flavor $\alpha$ using
\begin{equation}
    \kappa_\alpha n_\alpha^2 = \left.\frac{\partial n_\alpha}{\partial \mu}\right\rvert_T = -\frac{1}{m\omega^2 r} \frac{\partial n_\alpha}{\partial r}
\end{equation}

\begin{table*}
    \centering
    \begin{tabular}{||c c c c||} 
     \hline
     Figure & $(\ket{1}\ket{3},\ket{1}\ket{2},\ket{2}\ket{3})$ datasets & No. of samples, $n_b$ & Sample size, $N_b$ \\ [0.5ex] 
     \hline\hline
     2 & (45,45,45) & 100 & 40  \\
     \hline
     3C & (40,39,32) & 100 & 40  \\
     \hline
     3D & (682,858,863) & 200 & 700  \\
     \hline
     4 & (337,247,471) & 200 & 300  \\
     \hline
     5 & $\sim$(20,20,20) & 100 & 20  \\ 
     \hline
    \end{tabular}
 \caption{{\bfseries Bootstrap parameters used in the main text.}}
 \label{table: bootstrap datasets}
\end{table*}

\subsection{Experimental triplon density}
To access the triplon density $n^t=\langle n_1 n_2 n_3 \rangle$, we expand the singles density in terms of flavor densities, doublon densities, and triplon density as follows:
\begin{equation}
    n^s = n_1+n_2+n_3-2n^d_{13}-2n^d_{12}-2n^d_{23}+4n^t.
    \label{Eq: ns in terms of spins doublons and triplon}
\end{equation}
The singles density, $n_s$, is naturally measured during fluorescence imaging (fluorescence imaging leads to parity projection) without any additional splitting procedures, whereas the flavor and doublon densities can be obtained through the splitting procedure. We extract the triplon density by solving Eq.~\ref{Eq: ns in terms of spins doublons and triplon} for $n^t$ and estimate the error bars of $n^t$ by applying error propagation.

\subsection{Calculation of on-site pair correlations at unit filling for fixed number per flavor}
To obtain the theory curve in Fig.~5 of the main text, we need to find the chemical potentials at the trap center, $\mu_{0,\alpha}$, for fixed numbers of atoms per flavor, $N_\alpha$. First, we generate flavor densities as functions of $\mu_{0,\alpha}$, interactions, temperature, and trap radius using local density approximation in the atomic limit. The atom number for each flavor can be integrated from the density profiles. By imposing interactions and temperature as fixed parameters, we apply non-linear least squares fitting to extract $\mu_{0,\alpha}$. Since all parameters used in the HTSE are known, we calculate the on-site correlations and then select the values at unit filling.

\subsection{Error estimates for densities and correlations}
To estimate the statistical uncertainty of densities and correlations, we use a bootstrap analysis. We start with our original datasets and create $n_b$ bootstrap samples by randomly selecting $N_b$ subdatasets with replacements. For each of these samples, we calculate the mean of the densities and correlations. This process is repeated several times until we reach convergence. After these repetitions, we have a collection of means from the bootstrap samples. We then calculate the standard deviation of these means which gives us an estimate of the standard error, and we use them as error bars for our measurements \cite{efron1994introduction}. The number of bootstrap samples, $n_b$, and bootstrap sample size, $N_b$, are summarized in Table~\ref{table: bootstrap datasets}.

\begin{figure}
    \centering
    \includegraphics[width=0.6\linewidth]{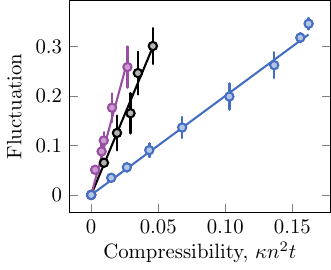}
    \caption{{\bfseries Thermometry using density fluctuations.} Black, violet, and blue correspond to the datasets shown in Figs.~ 2, 3A, and 3B of the main text. Dots represent experimental data. Solid lines are linear regressions, resulting in $k_BT/t=6.5(10)$, $10(2)$, and $1.8(2)$, respectively. Error bars are the standard error of the mean.}
    \label{fig:knnt vs fluc}
\end{figure}

\section{Thermometry}
To determine the temperature, $T$, of our three-flavor Fermi-Hubbard systems, we apply two approaches: we fit the spatial variation of the density of each flavor over the system size and we apply the fluctuation-dissipation theorem. Both methods result in consistent temperature within error bars.

\subsection{Thermometry using radial dependence of flavor density}
We fit the radial density profile of all three flavors simultaneously to either HTSE or NLCE by minimizing the $\chi^2$ defined by
\begin{equation}
    \chi^2(T,\mu_1,\mu_2,\mu_3) = \sum_{i}\sum_{j} \left(\frac{n_i(\mathbf{r}_j
    )-\bar{n}_i(\mathbf{r}_j)}{\bar{n}_i(\mathbf{r}_j)}\right)^2.
\end{equation}
Here, the index $i$ represents the flavor. We make use of the symmetry of the lattice and use the temperature, $T$, and the chemical potentials for each flavor at the trap center, $\mu_{0,i}$, as fit parameters.

We employ the local density approximation to determine the radial dependence of chemical potential from the trap center,
\begin{equation}
    \mu_i(\mathbf{r}) = \mu_{i,0}-\frac{1}{2}m\omega^2r^2,
\end{equation}
where $m$ is the atom mass and $\omega$ is the harmonic lattice confinement which is measured using a non-interacting Fermi gas (Sec.~\ref{Confinement calibration}).

\subsection{Thermometry using density fluctuation-dissipation theorem}
The density fluctuations can be used for thermometry which has been successfully demonstrated in a two-component mixture \cite{Hartke2020}. Here, we apply a similar approach to determine the temperature of our three-flavor mixture. The density fluctuation-dissipation theorem is given by

\begin{equation}
    \kappa n^2 = \frac{1}{k_B T} \sum_{\mathbf{a}} \Big(\expval{\hat{n}_\mathbf{r}\hat{n}_{\mathbf{r}+\mathbf{a}}}-\expval{\hat{n}_\mathbf{r}}\expval{\hat{n}_{\mathbf{r}+\mathbf{a}}}\Big),
\end{equation}
where $T$ is the temperature. $T$ can be obtained by performing a linear regression to $\kappa n^2$ and $\sum_\mathbf{a} (\dots)$ (Fig.~\ref{fig:knnt vs fluc}). We apply this technique to independently confirm our previous temperature measurement.

\end{document}